# An Energy-Efficient Accelerator Architecture with Serial Accumulation Dataflow for Deep CNNs

Mehdi Ahmadi, Shervin Vakili and J.M. Pierre Langlois
Department of Computer and Software Engineering
Polytechnique Montréal, Canada
{mehdi.ahmadi, shervin.vakili, pierre.langlois}@polymtl.ca

*Abstract*—Convolutional Neural Networks (CNNs) have shown outstanding accuracy for many vision tasks during recent years. When deploying CNNs on portable devices and embedded systems, however, the large number of parameters and computations result in long processing time and low battery life. An important factor in designing CNN hardware accelerators is to efficiently map the convolution computation onto hardware resources. In addition, to save battery life and reduce energy consumption, it is essential to reduce the number of DRAM accesses since DRAM consumes orders of magnitude more energy compared to other operations in hardware. In this paper, we propose an energy-efficient architecture which maximally utilizes its computational units for convolution operations while requiring a low number of DRAM accesses. The implementation results show that the proposed architecture performs one image recognition task using the VGGNet model with a latency of 393 ms and only 251.5 MB of DRAM accesses.

*Keywords— Energy-efficient design, convolutional neural networks (CNNs), deep learning, hardware accelerator*

## I. INTRODUCTION

Convolutional Neural Networks (CNNs) have attracted a lot of attention during past few years because of their high accuracy for recognizing images and detecting objects, even outperforming humans [1]-[3]. A CNN typically consists of many cascaded layers. The majority of the layers in a CNN are convolutional layers. It has been shown that increasing the number of convolutional layers can generally improve the image classification accuracy [3]. Increasing the number of layers, however, increases the computational complexity of CNNs and the number of parameters. For instance, VGGNet-16 [2], a well-known deep CNN, includes 13 convolutional layers and achieves a misclassification rate of 27% on the ImageNet dataset while requiring 14.7 M parameters and 30.6 G multiply-accumulation (MAC) operations for recognizing one image. A CNN is first trained using a training dataset, such as ImageNet. Then, the trained parameters are used in inference for recognizing the test images.

The outstanding advances in image recognition by CNNs motivates developers to deploy CNNs on portable and mobile devices. For mobile applications, however, CPUs are not fast enough to perform CNN inference in an acceptable amount of time. Therefore, in a mobile device, utilizing a hardware accelerator for convolutions can help to speed up CNN inference. Designers of CNN accelerators for mobile applications, however, face several implementation challenges regarding power and energy consumption. The power budget for such accelerators is typically in the range of a few hundreds milliwatts [4]. In addition, portable and mobile devices are normally battery-operated. This requires energy-efficient accelerator design to avoid rapid battery drain. From a hardware perspective, one off-chip DRAM access consumes orders of magnitude more energy compared to any on-chip operations [5]. On the other hand, using DRAM is unavoidable due to the large memory requirements of deep CNNs. Therefore, a low-energy accelerator should have a dataflow in which all the processing elements contribute in computations efficiently and the number of DRAM accesses is minimized.

During the past few years, several low-energy CNN accelerators have been developed on Application Specific Integrated Circuits (ASICs) [6]-[9]. Since convolutions typically account for more than 90% of the computations in a CNN [8], most of the accelerators are designed to perform the convolution operation efficiently. Moon *et al.* [7] proposed an accelerator called Envision to perform convolutional layers of AlexNet and VGGNet. In that design, the supply voltage and the precision of the computations are dynamically changed across the convolutional layer to save power and energy consumption. Chen *et al.* [8] proposed an NOC-like structure, called Eyeriss, to compute the convolutional layers of AlexNet and VGGNet. Eyeriss includes a large SRAM to store intermediate results and utilizes image batching to reduce the number of DRAM accesses. Ardakani *et al.* [9] proposed an architecture to compute the convolutional layers of VGGNet. They improved the energy efficiency by utilizing a dataflow inspired from the computational patterns of fully-connected (FC) layers.

In this paper, we propose a new architecture for accelerating computation of convolutional layers. A major advantage of this architecture is its capability to efficiently utilize all the computational units in all clock cycles. The proposed architecture includes several parallel units, each comprised of cascaded MAC operators whose outputs are accumulated in a serial fashion. We evaluate the architecture using the convolutional layers of VGGNet as a widely-used benchmark model.

## II. BACKGROUND

### A. Overview of CNN operation

Fig. 1 shows the convolution operation in a convolutional layer. A convolutional layer takes as input a 3-D matrix of size $IL \times IL \times IC$ and by convolving it with $M$ filters of size $FL \times FL \times FC$, it produces a 3-D outputs of size $OL \times OL \times OC$, where *IL*, *FL*, and *OL* denote the length and *IC*, *FC* and *OC* indicate the depth, also called the number of channels, of the corresponding matrices. Each element in the input or output matrix is called a feature and each filter element is called a weight. In the convolution, the number of input channels equals the number of filter channels, i.e., *IC=FC*. In addition, the number of filters is equal to the number of output channels, i.e., *M=OC*. The convolution operation is obtained as

$$y_k(m,n) = b^k + \sum_{c=0}^{IC-1}\sum_{j=0}^{FH-1}\sum_{i=0}^{FL-1} x_c(m\times s+j, n\times s+i)\times w_c^k(j,i)$$
$$0 \leq m,n \leq OL, 0 \leq k \leq OC, OL = (IL-FL+2z)/s+1 \quad (1)$$

where *y*, *b*, *x*, and *w* denote the elements in output, bias, input and filter matrices, respectively. *z* denotes the number of zero pads which are used to preserve the spatial size of the output features and *s* indicates the filter stride. In (1), for an element in a matrix, $x_c(r,q)$, *c* represents the channel index in the matrix, while *r* and *q* indicate the row and column of the element, respectively. Similarly, for the filter weights, $w_c^k(j,i)$, *c*, *j* and *i* denote the channel index, row number and column number, respectively, while *k* represents the filter index.

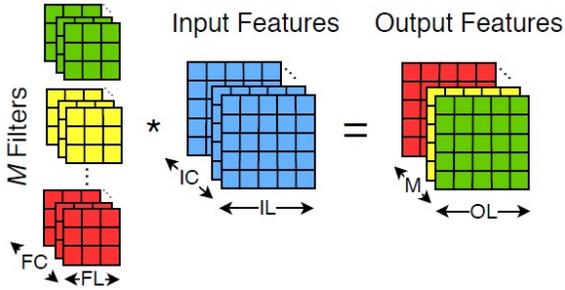

Fig. 1 Computation of a convolutional layer

### B. General functionality of CNN accelerators

A typical acceleration system consists of an inference engine, an off-chip DRAM, and a host processor [6]. When the execution reaches convolution operations, the processor uses the inference engine to speed up computations. Due to the large data requirement in deep CNNs, it is typical to use an off-chip DRAM to store the filter weights, input features and generated output features. The convolution process is started by fetching filter weights and input features from DRAM to the convolution inference engine. Inside the engine, there are several parallel units, also called Convolution Units (CU). Each unit consists of some Processing Elements (PE), each one equipped with a MAC operator, to perform dot product computations. In each clock cycle, partial results are generated by MACs and are stored in on-chip SRAMs. These partial results are accumulated and stored in the same memory locations in the next clock cycles. After completing the computations, the obtained output features are transferred from the on-chip SRAM to the off-chip DRAM. Due to massive data requirements and the limitations of SRAM size, the convolution engine cannot typically generate the entire output features of a convolutional layer in a single round of operation. In other words, the engine operates on a part of the input data to generate a part of the output features in each iteration. The engine works sequentially to generate the entire output features.

Properly partitioning of the filter and input feature matrices enables assignment of independent computations to parallel units, which consequently enhances the performance of the engine. A large group of existing architectures exploits inherent data-level parallelism in the convolutional layers by assigning the weights from each filter (output channel) to a distinct parallel unit [8]-[10]. The input features, on the other hand, can be shared among the parallel units through a global buffer [8] or pipelining [9]. In this paper, we propose a new architecture that falls into this group of parallel architectures and utilizes pipelined buffers.

### III. PROPOSED CNN ACCELERATOR

Fig. 2 shows a convolution engine with the proposed CU architecture. The engine consists of *U* parallel units, and each CU includes *N* multiply-accumulators (MACs) and *N* registers to keep the weight values. For the VGGNet accelerator, we set the value of *U* to 64 and *N* to 3.

Here, we explain the functionality of the first CU, i.e. CU #0, while the same process is performed in other parallel CUs which have the weights of other filters. The proposed architecture performs convolutions in a row-wise fashion. At the beginning of the convolution operation, the weights for the first filter row are fetched from DRAM and through the input IW, they are placed in registers WR0, WR1, and WR2. These weights are fixed in the registers, each of them provides an input of one multiplier. The other CU input is common to all three multipliers and receives the value of an input feature. The input feature is provided to each CU by pipeline registers through the input IX. In each clock cycle, a new input feature is fed to the pipeline registers while the previous input features advance one stage forward. When an input features arrives to the input IX of the CU, the dot product between filter weights and the input feature is performed by multipliers. The outputs of the multipliers are passed through multiplexers, added to the partial results from previous clock cycles and stored in accumulator registers, except for the last PE which writes the computed partial results directly into an SRAM. In other words, the PEs are connected together serially thorough accumulators and the content of accumulators are added from left to right until the result of three multiply and accumulations is stored in an SRAM by PE #2. The content of the SRAM can be fetched to PE #0 later for further processing.

As shown in Fig. 2, the proposed architecture utilizes duplicated SRAM units, i.e., SRAM M and SRAM P, and a ping-pong mechanism to enable overlap of the convolution computations with data transfer from the on-chip SRAMs to the off-chip DRAM. When the computations of a part of the output features are completed, the stored output features in an SRAM, e.g., SRAM M, have to be transferred to the DRAM. To prevent any data loss due to early overwriting, the PEs utilize another SRAM, e.g., SRAM P, to store the new partial results. In Fig. 2, the multiplexers M0 and M2 facilitate the handling of zero padding in the borders. These multiplexers replace the output of the multipliers by zeros in the borders.

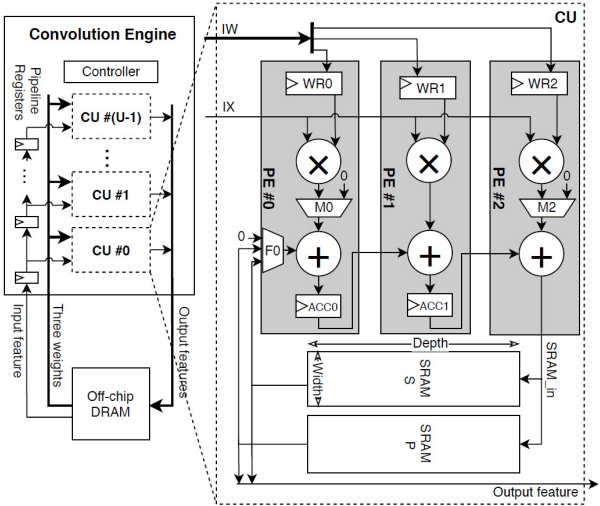

Fig. 2 A convolution engine with the proposed CU architecture.

Fig. 3 shows the deployed dataflow to perform 3×3 convolution in the proposed architecture using an example. This figure illustrates the register states and the accomplished operations in each clock cycle when convolving a filter row, comprised of three weights with three input features.

In clock cycle #1, the first input feature, $x_0(0,0)$, appears at the input of the CU while the weights of the first filter row, i.e., $w_0^0(0,0)$, $w_0^0(0,1)$, and $w_0^0(0,2)$, are stored in register WR0, WR1 and WR2, respectively. The dot product operation between the input feature and the filter weights generates the following values at the output of the multipliers: $x_0(0,0) \times w_0^0(0,0)$, $x_0(0,0) \times w_0^0(0,1)$, and $x_0(0,0) \times w_0^0(0,2)$. Since the last two values are not used in any subsequent computations they are discarded. In other words, $x_0(0,0) \times w_0^0(0,0)$ is stored in ACC0 while the ACC1 and SRAM are not loaded with new values.

In clock cycle #2, the new input feature, i.e., $x_0(0,1)$, appears at the input IX of the CU while the contents of weight registers, i.e., WR0, WR1 and WR2, remains unchanged. The second multiplier outputs $x_0(0,1) \times w_0^0(0,1)$, which is added to the content of ACC0 from the previous clock cycle. The result, i.e., $x_0(0,0) \times w_0^0(0,0) + x_0(0,1) \times w_0^0(0,1)$, is written into ACC1.

In clock cycle #3, the next input feature, $x_0(0,2)$, is given to the input of the CU. The output of the third multiplier, i.e., $x_0(0,2) \times w_0^0(0,2)$, is accumulated with the content of the ACC1 from previous clock cycle and produces $x_0(0,0) \times w_0^0(0,0) + x_0(0,0) \times w_0^0(0,0) + x_0(0,0) \times w_0^0(0,0)$. This is the result of convolving the first filter row with the first three input features and it is stored in an SRAM location.

This procedure continues uninterruptedly; in each cycle, a new partial result is produced in PE #2 and is stored into an SRAM. When the convolution of the first filter row with the corresponding input features is completed, the computations for the second filter row is started by placing its weights in registers WR0 to WR2. According to the dataflow, in each clock cycle a corresponding input feature appears at the input of the CU. When performing convolution for the second filter row, the generated partial results are accumulated with the corresponding ones generated by the first filter row. Therefore, the previously calculated partial results are fetched from SRAM, and through multiplexer F0, they are added to the newly generated partial results. This process is repeated for the last filter row and for the other filter channels. In the end, the corresponding output features are obtained and stored in the SRAM. The full SRAM then starts to transfer its contents to the off-chip DRAM. In the proposed architecture, during this SRAM-DRAM data transfer, the second SRAM is utilized by PEs for the new round of computational cycles. The procedure of generating output features in the CU #0 is performed in other CUs, in parallel, for the other filters.

Due to limited SRAM size, the convolution engine cannot normally generate all the output features in a single round of operation. Therefore, several iterations must be performed serially to complete the computation of the entire output features. In each iteration, all the filter weights are re-fetched from the off-chip DRAM. Increasing SRAM size enables storage of more partial results, and consequently increases the number of generated output rows in each iteration. This reduces the total number of required iterations that results in fewer re-fetches of weights. Increasing size of SRAMs, however, increases the on-chip cost and power consumption. Considering the trade-off between the number of DRAM accesses and the area and power consumption, we selected SRAMs of size 448 words in the proposed architecture.

IV. RESULTS AND DISCUSSIONS

The proposed architecture was modeled in Verilog HDL and was implemented in TSMC 65 nm LP CMOS technology at 200 MHz clock frequency using Cadence Genus. The word lengths of filter weights, input features and output features were set to 16 bits and the bit-width of SRAMs was set to 32 bits. TABLE I shows the implementation results for the proposed architecture and its comparison with the state of the arts. As shown in TABLE I, our architecture consumes 160 mW of power and 6.2 mm$^2$ of silicon area. In addition, it achieves a latency of 393 ms and requires 251.5 MB memory accesses to classify an image using VGGNet.

The FC-inspired architecture in [9] performs convolutions based on the computation patterns of fully connected (FC) layers. In that design, each MAC operator is considered as a neuron and its output is connected to an SRAM. The weights for the MAC operators are provided by a weight generator unit, which consists of five registers and two multiplexers. Using a separate SRAM unit to each MAC operator, however, introduces area and power consumption overhead compared to our design which uses unified SRAMs for all the PEs of each CU. In addition, the FC-inspired architecture was not actually implemented in an SRAM library from TSMC and the reported area and power consumption results were estimated. Furthermore, the paper has not described any mechanism to allow computations to proceed during SRAM-DRAM data transfers. In lack of such a mechanism, the convolution engine must be stalled for many clock cycles to first empty the contents of full SRAMs of all the parallel units and then it can start new computations. Our proposed architecture outperforms the FC-inspired architecture by 13.3% lower latency and by 24.1% fewer DRAM accesses.

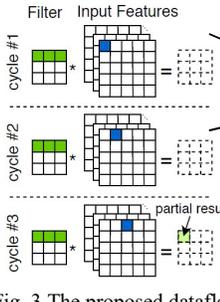

| cycle | IX | WR0 | WR1 | WR2 | ACC0 | ACC1 | SRAM_in (partial result) |
|---|---|---|---|---|---|---|---|
| 1 | $x_0(0,0)$ | $w_0^0(0,0)$ | $w_0^0(0,1)$ | $w_0^0(0,2)$ | $x_0(0,0) \times w_0^0(0,0)$ | 0 | 0 |
| 2 | $x_0(0,1)$ | $w_0^0(0,0)$ | $w_0^0(0,1)$ | $w_0^0(0,2)$ | $x_0(0,1) \times w_0^0(0,0)$ | $x_0(0,0) \times w_0^0(0,0) + x_0(0,1) \times w_0^0(0,1)$ | 0 |
| 3 | $x_0(0,2)$ | $w_0^0(0,0)$ | $w_0^0(0,1)$ | $w_0^0(0,2)$ | $x_0(0,2) \times w_0^0(0,0)$ | $x_0(0,1) \times w_0^0(0,0) + x_0(0,2) \times w_0^0(0,1)$ | $x_0(0,0) \times w_0^0(0,0) + x_0(0,1) \times w_0^0(0,1) + x_0(0,2) \times w_0^0(0,2)$ |
| 4 | $x_0(0,3)$ | $w_0^0(0,0)$ | $w_0^0(0,1)$ | $w_0^0(0,2)$ | $x_0(0,3) \times w_0^0(0,0)$ | $x_0(0,2) \times w_0^0(0,0) + x_0(0,3) \times w_0^0(0,1)$ | $x_0(0,1) \times w_0^0(0,0) + x_0(0,2) \times w_0^0(0,1) + x_0(0,3) \times w_0^0(0,2)$ |
| … | … | … | … | … | … | … | … |

Fig. 3 The proposed dataflow

Eyeriss [8] uses an NOC-based architecture utilizing a large global SRAM of size 108KB to buffer filter weights, input features and intermediate results. Eyeriss reduces the number of DRAM accesses by performing image classification on batches of three images instead of a single image. This results in a long computational latency of 4.3 seconds, which is prohibitive for many real-time applications. In addition, latency-sensitive applications normally require to process an image frame quickly and individually, before the next image frame arrives. Therefore, image batching is not a proper solution for a wide-range of real-time applications. Compared to Eyeriss, the proposed method has reduced the latency by 10.9× and the number of DRAM accesses by 21%.

Envision [7] is a CNN accelerator implemented in a smaller technology of 28 nm UTBB FD-SOI. From an architecture perspective, the MAC units can be dynamically configured to different word lengths for different convolutional layers according to the accuracy requirements of each layer. The proposed method outperforms Envision by offering 1.5× lower latency while utilizing 52% smaller area in terms of NAND2 gate counts. The number of DRAM accesses to perform image classification was not reported.

## V. CONCLUSION

In this work, we proposed an energy efficient architecture to perform convolutions in deep CNNs. The proposed architecture includes several parallel units in which the processing elements are connected together serially via accumulator registers. Therefore, the partial results are accumulated from the output of the first PE to the next PE until the last serial PE writes them to an SRAM. This dataflow maximally utilizes all the processing elements, which results in low-latency computations. In addition, properly selecting the SRAM size in the proposed architecture significantly decreased the number of DRAM accesses while introduced low hardware overheads. The evaluation of the proposed method on VGGNet showed that the proposed architecture outperforms the existing implementation on both the latency and the number of memory accesses. For example, the architecture achieves lower latency by 10.9× and requires 21% fewer DRAM accesses compared to Eyeriss when performing image classification.

TABLE I IMPLEMENTATION RESULTS

|  | FC-Inspired [9] | Envision [7] | Eyeriss [8] | Ours |
|---|---|---|---|---|
| Technology (nm) | 65 | 28 | 65 | 65 |
| On-chip SRAM (KB) | 86 | 144 | 181.5 | 224 |
| Frequency (MHz) | 200 | 200 | 200 | 200 |
| Bit precision | 16b | 1b..16b | 16b | 16b |
| #PEs | 192 | 256..1024 | 168 | 192 |
| Core Area (mm$^2$) | 3.5 | 1.87 | 12.25 | 6.2 |
| NAND2 Gate (K) | 1117 | 1950 | 1852 | 937 |
| Power (mW) | 260 | 26 | 236 | 140 |
| Latency (ms) | 453.3 | 598.8 | 4309.5 | 393.0 |
| Performance (Gops) | 67.7 | 51.3 | 21.4 | 78.1 |
| Efficiency (Gops/W) | 260.4 | 1973 | 90.7 | 557.9 |
| #DRAM access (MB) | 331.7 | NA | 321.1 | 251.5 |

ACKNOWLEDGMENT

This work was supported in part by the Natural Sciences and Engineering Research Council of Canada and by the Regroupement Stratégique en Microsystèmes du Québec.